\documentclass[showpacs,12pt,preprint,preprintnumbers,nofootinbib,groupedaddress,superscriptaddress,amsmath,amssymb]{revtex4}

\usepackage{graphicx}
\usepackage{dcolumn}
\usepackage{bm}
\usepackage{amssymb}
\usepackage{amsmath}
\usepackage{epsfig} 

\def\beq{\begin{equation}}
\def\eeq{\end{equation}}
\def\bea{\begin{array}}
\def\eea{\end{array}}
\def\be{\begin{equation}}
\def\ee{\end{equation}}
\def\ba{\begin{eqnarray}}
\def\ea{\end{eqnarray}}

\def\to{\rightarrow}

\def\[{\left[}
\def\]{\right]}
\def\({\left(}
\def\){\right)}

\def\sm0{{\widetilde{m}_0}}

\def\U1em{{U(1)_{\rm em}}}
\def\to{\rightarrow}

\def\sq2{\sqrt{2}}

\def\ee{e^+e^-}

\def\End{\end{document}}

\newcommand{\gsim}{\mbox{ \raisebox{-1.0ex}{$\stackrel{\textstyle >}
{\textstyle \sim}$ }}}
\newcommand{\lsim}{\mbox{ \raisebox{-1.0ex}{$\stackrel{\textstyle <}
{\textstyle \sim}$ }}}

\def\fsl#1{\setbox0=\hbox{$#1$}                 
   \dimen0=\wd0                                 
   \setbox1=\hbox{/} \dimen1=\wd1               
   \ifdim\dimen0>\dimen1                        
      \rlap{\hbox to \dimen0{\hfil/\hfil}}      
      #1                                        
   \else                                        
      \rlap{\hbox to \dimen1{\hfil$#1$\hfil}}   
      /                                         
   \fi}

\begin{document}

\title{ Multi-Higgs portal dark matter under the CDMS II results}

\author{Mayumi, Aoki}
\email{mayumi@tuhep.phys.tohoku.ac.jp}
\affiliation{Department~of~Physics,~Tohoku~University,~Aramaki,~Aoba,~Sendai,~Miyagi~980-8578,~Japan}
\author{Shinya Kanemura}
\email{kanemu@sci.u-toyama.ac.jp}
\affiliation{Department~of Physics,~University~of~Toyama,~3190~Gofuku,~Toyama~930-8555,~Japan}
\author{Osamu Seto}
\email{osamu@hgu.jp}
\affiliation{Department of Architecture and Building Engineering, Hokkai-Gakuen University,
 Sapporo 062-8605, Japan}
%
\preprint{TU-860, UT-HET 033, HGU-CAP 001}
\pacs{\, } 

\begin{abstract}
In a scenario of Higgs portal dark matter, Higgs exchange processes are
 essential for both dark matter annihilation in the early Universe
 and direct search experiments.
The CDMS II collaboration has recently released their final results on direct dark matter searches.
We study a scalar dark matter model with multi-Higgs doublets under the
 constraint from the CDMS II results and also from the WMAP data.
We find that the possible maximal value for the branching ratio
of the invisible decay of the Higgs boson can be significantly greater than 
that in the Higgs portal model with one Higgs doublet, in particular,
 for the case of the so-called Type-X Yukawa interaction. 
 Therefore, the search for the invisible decay of the Higgs boson
at the CERN Large Hadron Collider and future collider experiments 
would provide useful information not only for
the nature of dark matter but also for the structure of 
the Higgs sector even without directly detecting any extra scalar boson.   
 \pacs{\, 
\hfill   ~~ [\today] }
\end{abstract}

\maketitle

\setcounter{footnote}{0}
\renewcommand{\thefootnote}{\arabic{footnote}}

\section{Introduction}

Various astrophysical and cosmological observations provide
evidence of the existence of dark matter (DM)~\cite{WMAP}.
The most interesting and promising candidate for DM is 
weakly interacting massive particles (WIMPs).
WIMP DM is directly detectable.
Many DM direct search experiments are operating and planed.
For example, 
the ongoing experiments are DAMA/LIBRA~\cite{Bernabei:2008yi}, 
EDELWEISS II~\cite{EDEL}, ZEPLIN II~\cite{ZEPLIN-II}, and  
XENON 10~\cite{XENON10}. Recently, CDMS II~\cite{CDMSII} has just finished,
while XENON 1T, superCDMS, and XMASS~\cite{XMASS} are planed.
Direct DM searches have been done by looking for 
the elastic scattering of WIMP with target nuclei 
through nuclear recoil. 
Higgs-boson-exchange processes mainly lead to 
scalar (in other words, spin-independent (SM)) couplings
between nuclei and WIMP.
Hence, the structure of a Higgs sector and its coupling with DM
are crucial for direct DM searches.

Among WIMP DM candidates, 
a class of models is categorized as ``Higgs portal dark matter'', 
in which a DM particle interacts with Standard Model (SM) particles
through only Higgs exchange processes.
The minimal model was constructed by adding only one 
new $Z_2$ parity-odd real scalar field to the SM~\cite{McDonald:1993ex,Burgess:2000yq}.
A variety of such models has been proposed~\cite{Deshpande:1977rw,Ma:2006km,Barbieri:2006dq,Aoki:2008av,Goh:2009wg}.
Some of them are motivated in the context of radiative seesaw models~\cite{Ma:2006km,Aoki:2008av}.

A remarkable but common feature of Higgs portal DM models 
is the invisible decay of the Higgs boson due 
to Higgs-DM couplings for the case that the DM mass is smaller than 
one half of the Higgs boson mass~\cite{Bento:2000ah,Bento:2001yk}.
In the minimal Higgs portal DM model, 
the upper bound is obtained for the branching ratio
of the invisible decay from the new CDMS II results~\cite{He:2009yd,Farina:2009ez,Kadastik:2009gx,Cheung:2009wb}.

In this Letter, we study a $Z_2$-odd scalar Higgs portal DM scenario in the
framework of multi-Higgs doublet models.
Such a scenario can appear, for example, in the  effective theory of the
three-loop-induced neutrino mass model~\cite{Aoki:2008av}, in which not
only tiny neutrino masses but also DM as well as baryon asymmetry may 
be explained simultaneously by the TeV scale physics.
The upper bound on the branching ratio of the Higgs boson invisible
decay is evaluated in the model with two Higgs doublets and a real $Z_2$-odd
singlet scalar field under the CDMS II results. 
In the analysis, a specific Yukawa interaction (the Type-X Yukawa
interaction~\cite{Goh:2009wg,typeX,Su:2009fz,Logan:2009uf}) is employed,
which is used in Ref.~\cite{Aoki:2008av} and is defined under the other
(softly broken) discrete symmetry ($\tilde{Z}_2$) for avoiding flavor
changing neutral current (FCNC).
We also give a comment on the results assuming the other types of Yukawa
interaction, such as so-called Type-II.
We then discuss the difference of the upper bound from that in the
minimal model with one Higgs doublet.
We show exclusive features of scalar Higgs portal DM with multi-Higgs doublets.

\section{Model}

We consider the model in which two Higgs doublet fields $\Phi_1$ and $\Phi_2$
and one real singlet scalar field $\eta$ are included.
A discrete $Z_2$ symmetry is introduced in the model, and the odd charge 
is assigned for $\eta$ to guarantee the stability as a candidate of DM.
The scalar potential is given by\footnote{This potential has been
studied in the different context in Ref.~\cite{Grzadkowski:2009iz}.} 
\begin{eqnarray}
  V = \frac{1}{2}\mu_{\eta}^2\eta^2 + \lambda_{\eta}\eta^4
       + \sum_{i=1,2} \sigma_i|\Phi_i|^2\eta^2 + V(\Phi_1, \Phi_2), \label{eq:pot}
\end{eqnarray}
where $\mu_{\eta}^{2}$ is the invariant squared mass of $\eta$, and
$V(\Phi_1,\Phi_2)$ is the potential of the two Higgs doublet model. 
We neglect the CP violating phase, so that all the coupling constants
are real. 
After electroweak symmetry breaking, neutral component fields in the Higgs doublets 
are parameterized as
\begin{equation}
 \phi_i^0 = \frac{1}{\sqrt{2}}(v_i+h_i +i z_i), \,\,\, (i=1,2),
\end{equation}
where $v_i$ are the vacuum expectation values (VEVs) that satisfy $v_1^2+v_2^2=v^2=(246 {\rm GeV})^2$
and $\tan\beta = v_2/v_1$.
The mass matrix for $h_1$ and $h_2$ is diagonalized by introducing
the mixing angle $\alpha$, and  two CP-even states $h$ and $H$ are the
mass eigenstates of the CP-even bosons.
The CP-odd scalar bosons $z_1$ and $z_2$ mix with each other,
and becomes the CP-odd Higgs $A$ and the longitudinal mode of the $Z$ boson.
In total, from $\Phi_1$ and $\Phi_2$
five physical states  appear;
i.e., two CP-even ($h, H$), one CP-odd ($A$), and charged ($H^{\pm}$)
scalar bosons. 

In the limit of $\sin(\beta-\alpha)=1$, $h$ is the SM-like Higgs boson;
i.e, all the coupling constants with SM fields coincide with those
of the SM Higgs boson at the tree level~\cite{Gunion:2002zf}.
On the other hand, $H$ does not receive the VEV in this limit. 
In this Letter, we always take this limit (the SM-like limit) for simplicity.
The mass of the SM-like Higgs boson $h$ is bounded from below
($m_h > 114$ GeV) from the LEP experiment, while that of $H$ 
can be lower than 100 GeV because it does not couple to the weak gauge
bosons in this limit.

Multi-Higgs doublet models in general suffer from dangerous FCNC.
To avoid FCNC, we impose a softly broken discrete symmetry $\tilde{Z}_2$
under the transformation $\Phi_1 \to \Phi_1$ and $\Phi_2 \to - \Phi_2$.
\begin{table}[tb]
\begin{center}
\begin{tabular}{|c||c|c|c|c|c|c|c|c|c|}
\hline
& $\xi_h^u$ & $\xi_h^d$ & $\xi_h^\ell$
& $\xi_H^u$ & $\xi_H^d$ & $\xi_H^\ell$
& $\xi_A^u$ & $\xi_A^d$ & $\xi_A^\ell$ \\ \hline
Type-I
& $c_\alpha/s_\beta$ & $c_\alpha/s_\beta$ & $c_\alpha/s_\beta$
& $s_\alpha/s_\beta$ & $s_\alpha/s_\beta$ & $s_\alpha/s_\beta$
& $\cot\beta$ & $-\cot\beta$ & $-\cot\beta$ \\
Type-II
& $c_\alpha/s_\beta$ & $-s_\alpha/c_\beta$ & $-s_\alpha/c_\beta$
& $s_\alpha/s_\beta$ & $c_\alpha/c_\beta$ & $c_\alpha/c_\beta$
& $\cot\beta$ & $\tan\beta$ & $\tan\beta$ \\
Type-X
& $c_\alpha/s_\beta$ & $c_\alpha/s_\beta$ & $-s_\alpha/c_\beta$
& $s_\alpha/s_\beta$ & $s_\alpha/s_\beta$ & $c_\alpha/c_\beta$
& $\cot\beta$ & $-\cot\beta$ & $\tan\beta$ \\
Type-Y
& $c_\alpha/s_\beta$ & $-s_\alpha/c_\beta$ & $c_\alpha/s_\beta$
& $s_\alpha/s_\beta$ & $c_\alpha/c_\beta$ & $s_\alpha/s_\beta$
& $\cot\beta$ & $\tan\beta$ & $-\cot\beta$ \\
\hline
\end{tabular}
\end{center}
\caption{The mixing factors in Yukawa interactions in Eq.~\eqref{Eq:Yukawa}} \label{Tab:MixFactor}
\end{table}
The Yukawa interactions are expressed in terms of mass eigenstates of the Higgs bosons as
\begin{align}
{\mathcal L}_\text{yukawa}^\text{THDM} =
&-\sum_{f=u,d,\ell} \( \frac{m_f}{v}\xi_h^f{\overline
f}fh+\frac{m_f}{v}\xi_H^f{\overline
f}fH-i\frac{m_f}{v}\xi_A^f{\overline f}\gamma_5fA\)\nonumber\\
&-\left\{\frac{\sqrt2V_{ud}}{v}\overline{u}
\left(m_u\xi_A^u\text{P}_L+m_d\xi_A^d\text{P}_R\right)d\,H^+
+\frac{\sqrt2m_\ell\xi_A^\ell}{v}\overline{\nu_L^{}}\ell_R^{}H^+
+\text{H.c.}\right\},\label{Eq:Yukawa}
\end{align}
where $P_{L/R}$ are projection operators for left-/right-handed fermions,
and the factors $\xi^f_\varphi$ are listed in TABLE~\ref{Tab:MixFactor}.
There are four ways of charge assignment under this $\tilde{Z}_2$ parity, 
thus correspondingly four independent types of Yukawa interaction are possible~\cite{Barger:1989fj,Grossman:1994jb}.
The typical example of so-called Type-II Yukawa interactions is 
that of the minimal supersymmetric standard model.
The Type-X Yukawa interaction~\cite{Goh:2009wg,typeX,Su:2009fz,Logan:2009uf},
where one of the Higgs doublet couples to only quarks and the other does
to only leptons, is adopted in the model for
radiative generation of tiny neutrino masses with including
the scalar DM proposed in Ref.~\cite{Aoki:2008av},   
whose Higgs sector contains two Higgs doublets
and a DM candidate $Z_2$-odd singlet scalar field as well as
some heavier particles.
Therefore, our present model given in Eq.~(\ref{eq:pot}) can be
regarded as the effective theory of the model in Ref.~\cite{Aoki:2008av}.
Thus, in this Letter, we mainly study the model with the Type-X Yukawa
interaction, and then give a short comment on the cases of the other
types for Yukawa interactions. 

Even in the SM-like limit, the total decay width of the SM-like Higgs boson
$h$ in our model can drastically change from the SM value
 when $m_\eta < m_h/2$ because of the additional invisible $h\to \eta\eta$ decay.  The total width of $h$ is given by
\begin{equation} 
\Gamma_{\rm tot} = \Gamma_{\rm vis} + \Gamma_{\rm inv},
\end{equation}
where $\Gamma_{\rm vis}$ denotes the width for 
Higgs boson decays into SM particle contents. 

In the SM-like limit, $\Gamma_{\rm vis}$ in our model coincides with that in the SM at the lowest order. 
The invisible decay width $\Gamma_{\rm inv}$ of the SM-like Higgs boson is
computed as 
\begin{eqnarray}
 \Gamma_{\rm inv}(h \rightarrow \eta\eta) = \frac{v^2}{32 \pi m_h}\sqrt{1-\frac{4 m_{\eta}^2}{m_h^2}}
 \left| -\sigma_1 \sin\alpha\cos\beta +\sigma_2 \cos\alpha\sin\beta \right|^2 .\label{GammaInv}
\end{eqnarray}
The corresponding formula in the minimal Higgs portal DM model with a scalar
doublet $\Phi$ and a real scalar field $\eta$ is obtained from 
Eq.~(\ref{GammaInv}) by replacing $(-\sigma_1\sin\alpha\cos\beta+\sigma_2\cos\alpha\sin\beta)$
by $2\sigma_{\rm m}$  
when the DM-Higgs coupling is given by ${\cal L}_{\rm int} =
\cdot\cdot\cdot -\sigma_{\rm m} \eta^2 |\Phi|^2 +\cdot\cdot\cdot$.
The branching ratio for the invisible decay is given by
\begin{equation}
   B_{\rm inv}(h\to\eta\eta) \equiv   \frac{ \Gamma_{\rm inv}}{ \Gamma_{\rm tot} }.\label{BInv}
\end{equation}

\section{Upper bound on the invisible decay branching ratio}

Now we consider the invisible decay branching ratio
of the SM-like Higgs boson $h$.
The size is proportional to the square of the $h\eta\eta$ coupling, 
but constrained by two issues.

One is from the direct DM search,
the most stringent bound comes from the latest CDMS II results for the 
relatively large mass region while 
the constraint from XENON 10 is slightly stronger for smaller mass values.
A too large coupling conflicts with the
fact that CDMS has just  observed only two possible events 
and others have obtained null results until now.
The DM SI cross section for a proton is given as
\begin{equation}
\sigma_p^{SI} =  \frac{m_p^2}{\pi (m_{\eta}+m_p)^2}f_p^2,   
\end{equation}
with
\begin{eqnarray}
\frac{f_p}{m_p} = \left(\sum_{q=u,d,s} f_{T\,q}^{(p)}
   + \frac{2}{27}\sum_{q=c,b,t} f_{TG\,q}^{(p)} \right)\frac{f_q}{m_q},
\end{eqnarray}
where $m_p$ is the proton mass and 
$f_p$ is the effective coupling with proton and 
$f_q^{(p)}$ is the hadronic matrix elements.
The effective coupling $f_q$ with a quark is model-dependent.
In the model in Eq.~(\ref{eq:pot}) with the Type-X Yukawa coupling, this
is calculated at the tree level as 
\begin{eqnarray}
 \frac{f_q}{m_q} &=& \frac{(-\sigma_1\sin\alpha\cos\beta+\sigma_2\cos\alpha\sin\beta)}{2m_h^2}
\frac{\cos\alpha}{\sin\beta} 
+\frac{(\sigma_1\cos\alpha\cos\beta+\sigma_2\sin\alpha\sin\beta)}{2m_H^2}\frac{\sin\alpha}{\sin\beta}.
\end{eqnarray}
In the minimal Higgs portal DM model, it is given by $f_q/m_q = \sigma_{\rm m}/m_h^2$.

The other is the cosmological DM abundance determined by thermal freeze out.
The relic mass density is evaluated as
\begin{equation}
\Omega_{\eta} h^2 =
 1.1 \times 10^9 \frac{m_{\eta}/T_d}{\sqrt{g_*}M_P\langle\sigma v\rangle}
  {\rm GeV^{-1}} ,
\end{equation}
with the Planck mass $M_P$, the total number of 
relativistic degrees of freedom in the thermal bath $g_*$, and 
the decoupling temperature $T_d$.
For the Type-X Yukawa interaction, the processes of
$\eta\eta\to b\bar b$ and $\eta\eta\to\tau^+\tau^-$ are dominant
when $m_\eta < m_W^{}$, and  
the thermal averaged product of annihilation cross section and 
relative velocity is evaluated as~\cite{Aoki:2008av} 
\begin{eqnarray}
\langle \sigma v \rangle
 &\simeq& 
 \frac{s}{16\pi m_{\eta}^2}
 \left. \left[ 
 3m_b^2
 \left| \frac{-\sigma_1 \sin\alpha\cos\beta +\sigma_2
  \cos\alpha\sin\beta}{s-m_h^2+i m_h\Gamma_{\rm tot}^h}
  \left(\frac{\cos\alpha}{\sin\beta}\right) \right.\right.\right. \nonumber\\
&& \left.\left.\left.\hspace*{3cm} +\frac{\sigma_1 \cos\alpha\cos\beta+\sigma_2
 \sin\alpha\sin\beta}{s-m_H^2+im_H\Gamma_{\rm tot}^H}
 \left(\frac{\sin\alpha}{\sin\beta}\right)
 \right|^2 \right.\right. \nonumber\\
&& \left.\left. \hspace*{1.4cm}+ m_{\tau}^2
 \left| \frac{-\sigma_1 \sin\alpha\cos\beta +\sigma_2 \cos\alpha\sin\beta}{s-m_h^2+i m_h\Gamma_{\rm tot}^h}
   \left(\frac{-\sin\alpha}{\cos\beta}\right)\right.\right.\right. \nonumber\\
&&\left.\left.\left.\hspace*{3cm}+\frac{\sigma_1 \cos\alpha\cos\beta+\sigma_2
 \sin\alpha\sin\beta}{s-m_H^2+im_H\Gamma_{\rm tot}^H}
\left(\frac{\cos\alpha}{\cos\beta}\right) \right|^2
  \right] \right|_{s=4m_{\eta}^2}, 
\end{eqnarray}
where $\Gamma_{\rm tot}^H$ is the total width of $H$.
In the minimal Higgs portal DM model, it is given as $\langle \sigma v
\rangle \sim 3 s/(4\pi m_\eta^2) \, |\sigma_{\rm m}/(s-m_h^2+i m_h
\Gamma_{\rm tot}^h)|^2$ with $s \simeq 4m_{\eta}^2$.   
Too large (small) coupling constants $\sigma_i$ 
correspond to the over-annihilation (over-abundance) of DM.
We evaluate the consistent parameter region of the $h\eta\eta$ coupling
and DM mass. 

We examine the upper bound on $B_{\rm inv}(h\to\eta\eta)$ from the CDMS II and the
XENON 10 for some parameter sets in the case of the Type-X interaction.
As stated, we work in the SM-like limit $\sin(\beta-\alpha)=1$.  
We use the average of the coupling constants $\sigma \equiv
(\sigma_1+\sigma_2)/2$
to show the typical scale of couplings and the difference
$\Delta \sigma \equiv \sigma_1-\sigma_2$ to see the effect of the difference
instead of $\sigma_1$ and $\sigma_2$.
For the numerical evaluation, we here show the results in the following two
simple cases with $\tan\beta=1$ (Set A) and $\tan\beta=10$ (Set B).
The mass of the SM-like Higgs boson $h$ is set to be $m_h=120$ GeV.
The other input parameters are commonly taken as $m_H^{}=90$ GeV and $\Delta\sigma=0.02$.
These parameter sets are not excluded by the current data.
We note that Set B approximately corresponds to the scenario discussed
in Ref.~\cite{Aoki:2008av} in the context of successful radiative seesaw
scenario with satisfying the constraint from dark matter abundance
and the condition for strongly first order phase transition for
electroweak baryogenesis.

\begin{figure}[t]
\begin{minipage}{0.49\hsize}
\includegraphics[width=8cm]{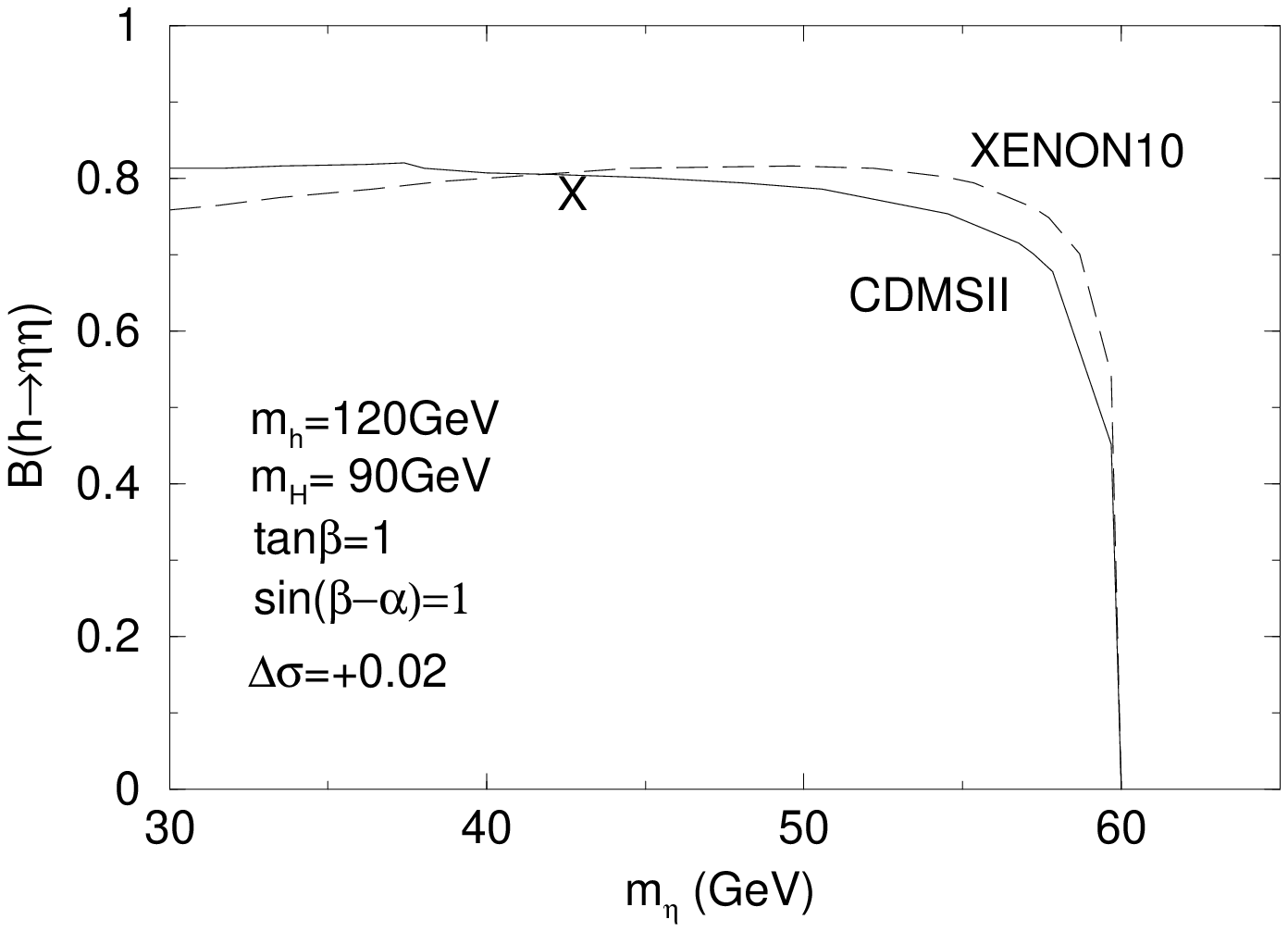}
\end{minipage}
\begin{minipage}{0.49\hsize}
\includegraphics[width=8cm]{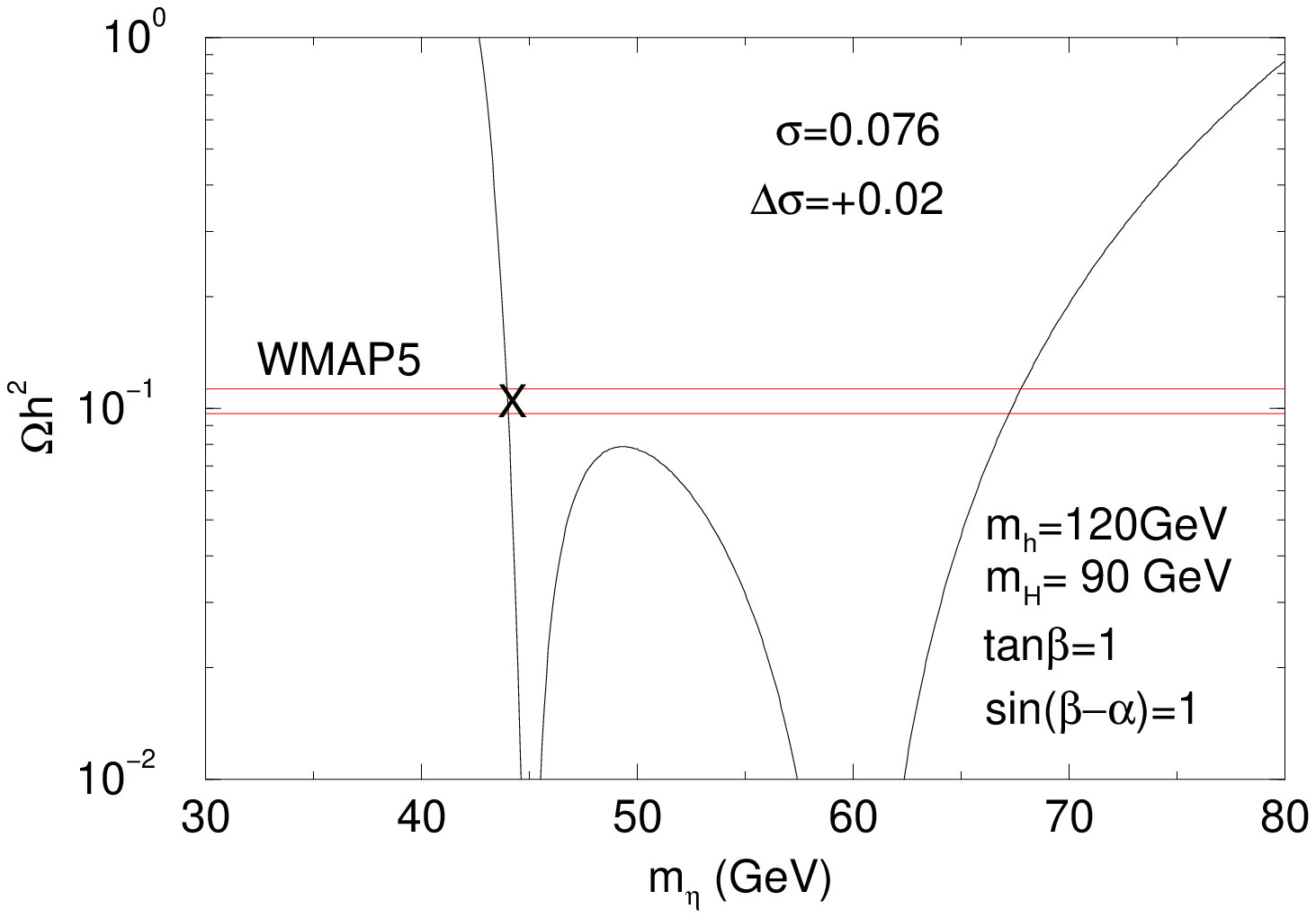}
\end{minipage}
  \caption{
 [Left] The constraint on the decay branching ratio $B_{\rm inv}(h\to \eta\eta)$
 for the invisible decay of the SM-like Higgs boson into a DM pair from
 the CDMS II results and the XENON 10 results in Set A.
[Right]
 The thermal abundance $\Omega h^2$ of DM as a function of the DM mass
 in Set A with $\sigma=0.076$. }
  \label{fig:setA}
\end{figure}

\begin{figure}[t]
\begin{minipage}{0.49\hsize}
\includegraphics[width=8cm]{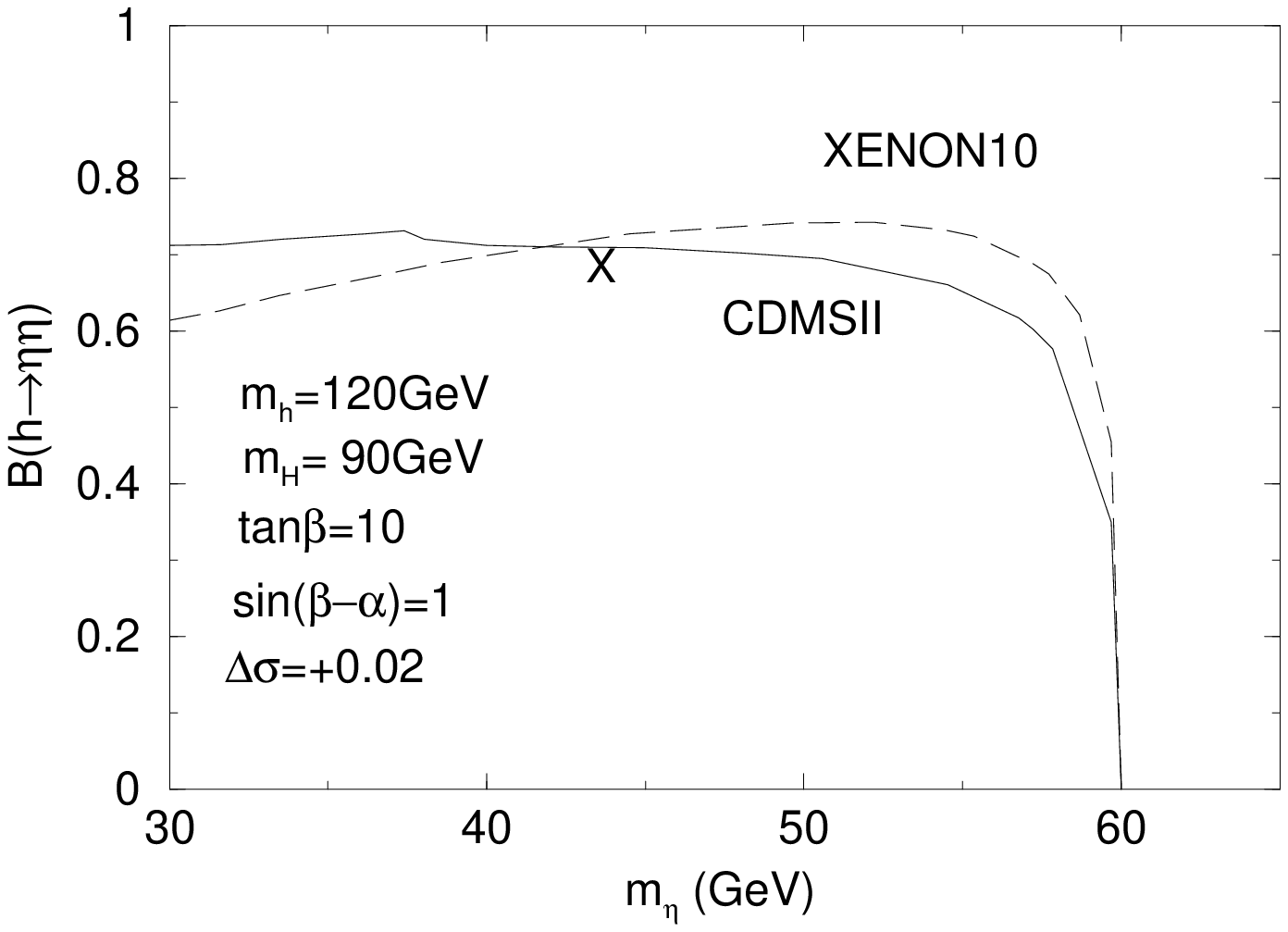}
\end{minipage}
\begin{minipage}{0.49\hsize}
\includegraphics[width=8cm]{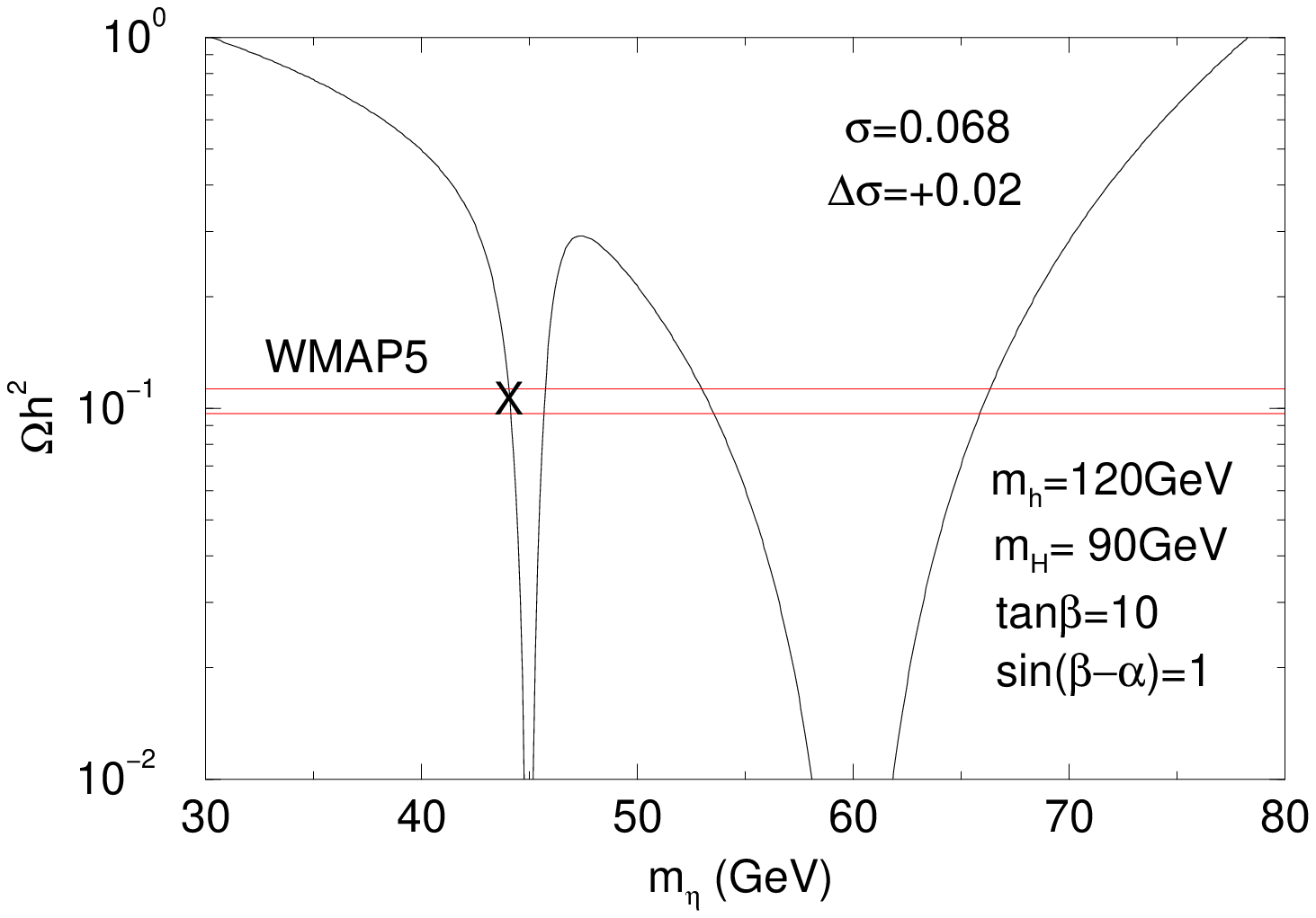}
\end{minipage}
  \caption{
[Left] The constraint on the decay branching ratio $B_{\rm inv}(h\to \eta\eta)$
 for the invisible decay of the SM-like Higgs boson into a DM pair from
 the CDMS II results and the XENON 10 results in Set B.
[Right]
 The thermal abundance $\Omega h^2$ of DM as a function of the DM mass
 in Set B with $\sigma=0.068$. }
  \label{fig:setB}
\end{figure}

The left panel of Fig.~\ref{fig:setA} shows the constraint on 
the Higgs invisible decay branching ratio from direct DM searches 
in the model with the Type-X Yukawa interaction and
$\sin(\beta-\alpha)=1$ for Set A 
 $(m_h, m_H,\Delta\sigma, \tan\beta) = (120~{\rm GeV}, 90~{\rm GeV}, 0.02, 1)$.
The upper bound on $B_{\rm inv}(h \to \eta\eta)$ does not depend on the DM mass much
 and about $B_{\rm inv}(h \to \eta\eta) \sim 0.8$ is allowed for $m_{\eta} \lesssim 50$ GeV,
while the bound becomes stringent for $m_{\eta} > 55$ GeV.
Around $m_{\eta} \simeq 43$ GeV (near the $H$-resonance), 
we can obtain the maximal value of $B_{\rm inv}(h \to \eta\eta)\simeq 0.8$ which corresponds to
$\sigma \simeq 0.076$ through Eqs.~(\ref{GammaInv}) and (\ref{BInv}). 
The right figure shows the point
 $(m_{\eta}, \sigma) \simeq (43~{\rm GeV}, 0.076)$ with the same other parameters 
indeed satisfies the WMAP  constraint $\Omega h^2 \simeq 0.1$.

The left panel of Fig.~\ref{fig:setB} similarly 
shows the constraint on the Higgs invisible decay branching ratio 
from direct DM searches in the model with the Type-X Yukawa interaction and
$\sin(\beta-\alpha)=1$ for Set B
 $(m_h, m_H, \Delta\sigma, \tan\beta) = (120~{\rm GeV}, 90~{\rm GeV}, 0.02, 10)$.
As compared to Set A shown in Fig.~~\ref{fig:setA},
the bound on $B_{\rm inv}(h \to \eta\eta)$ becomes stringent.
However, a large $B_{\rm inv}(h \to\eta\eta) \simeq 0.7$ is still realized  
for $m_{\eta} \simeq 43$ GeV  (near the $H$-resonance) and the large invisible width
is obtained for $\sigma \simeq 0.068$. This point satisfies 
the WMAP constraint on the DM abundance as shown in the right figure.

We have observed that there are the parameter sets in our model with the
Type-X Yukawa interaction and $\sin(\beta-\alpha)=1$, in which
a maximal values of $B_{\rm inv}(h\to\eta\eta)$ suggested by the direct search
results are consistent with the WMAP results.
In Set A (Set B), $B_{\rm inv}(h\to\eta\eta)=0.8$ $(0.7)$ can be realized for
$m_h=120$ GeV and $m_\eta\simeq 43$ GeV around the edge of the resonance of $H$.
On the other hand, in the model with the minimal Higgs portal DM model,
we obtain 
$B_{\rm inv}(h\to\eta\eta) \lsim 0.63$ for the same value of $m_h$ but
for $m_\eta \sim 55$ GeV (near the $h$ resonance) in the same calculation manner\footnote{
In Refs.~\cite{He:2009yd} and \cite{Farina:2009ez}, somewhat larger values are 
reported for the upper bound of $B_{\rm inv}(h\to\eta\eta)$ in the minimal
Higgs portal DM model.
The difference between our result and their results
mainly comes from the different choice for the values of the hadronic matrix
elements. In our analysis, the values in Ref.~\cite{Ellis:2008hf} are 
consistently used.}.
Therefore, if $B_{\rm inv}(h\to\eta\eta) \gg 0.63$ will be measured at the LHC, it
will indicate a non-minimal Higgs sector in the Higgs portal DM scenario
even when no extra Higgs boson will be found there  yet. 

The invisible decay of the Higgs boson can be detected at the CERN Large
 Hadron Collider (LHC) 
 if $B_{\rm inv}(h\to\eta\eta) > 0.25$~\cite{INV_LHC}. At the International Linear Collider
 (ILC), invisible decays of the SM-like Higgs boson $h$ can be tested
 when $B_{\rm inv}(h\to\eta\eta) > $ only a few \%~\cite{INV_ILC}. Therefore, we can distinguish
the maximal value of the branching ratio evaluated in our model
from the upper bound in the minimal one doublet model with a $Z_2$-odd singlet scalar boson.

As we have seen, the extra scalar boson $H$ has to be lighter than the SM-like
Higgs boson $h$ in order to obtain $B_{\rm inv}(h\to\eta\eta) \gsim  0.63$. 
Phenomenology of extra Higgs bosons in the Type-X two Higgs doublet
model has been studied in Ref.~\cite{Goh:2009wg,typeX,Su:2009fz,Logan:2009uf,Belyaev:2009zd}. 
At the LHC such a light $H$ ($m_H \sim 90$ GeV) can be produced via
gluon fusion processes $gg\to H$ and $gg\to g H$, and also 
the Drell-Yan type processes $q\bar{q}' \to HH^+$ and $q\bar{q} \to AH$.
The decay pattern of $H$ largely depends on $\tan\beta$.
For $\tan\beta\sim 1$, it decays mainly into $bb$, while for $\tan\beta \sim 10$ the 
leptonic decay modes into $\tau^+\tau^-$ and $\mu^+\mu^-$ are dominant.
If such a light $H$ is identified and large $B_{\rm inv}(h\to\eta\eta) \gsim 0.63$
is confirmed at the LHC, then the two Higgs portal DM scenario can be tested.
On the other hand, if only $B_{\rm inv}(h\to\eta\eta) \gg 0.63$ is
measured at the LHC without detecting $H$, then we could
obtain indirect information on the extended Higgs sector in the Higgs
portal DM scenario before direct detection of the extra Higgs bosons. 

In this Letter, we have studied only two Higgs doublet model 
with Type-X Yukawa coupling.
However, relaxation of the upper bound on the invisible decay branching ratio 
seems to be generic for other multi-Higgs doublet models as
well\footnote{
It is easily understood from Eq.~(\ref{Eq:Yukawa}) that for $\sin(\beta-\alpha)=1$ 
and $\tan\beta=1$ the visible width of the extra Higgs boson $H$ is independent of 
the types of Yukawa interaction~\cite{typeX}, and the abundance of $\eta$ is also calculated to 
be almost common when the mass of $\eta$ to be near the $H$ and $h$ resonances.  
Therefore, our result of $B_{\rm inv}(h \to\eta\eta) \lsim 0.8$ for Set A is essentially 
independent of the type of Yukawa interaction.}.
The essence of this enhancement comes from the fact that
the relevant interaction of DM for direct DM search experiments is 
both $h$-mediation and $H$-mediation, while
only the coupling with $h$ is relevant to 
the invisible decay of the SM-like Higgs boson.
Detailed study will be shown elsewhere~\cite{aks_cdmsfull}.

\section{Conclusions}

We studied the branching ratio of the Higgs invisible decay in the  model with 
multi-Higgs doublets and one scalar singlet DM field, mainly assuming
the Type-X Yukawa interaction.
We could rewrite 
the latest CDMS II and XENON 10 excluded region 
into an upper bound of Higgs invisible decay for a given parameter set.
If the two suspicious CDMS events are indeed due to the WIMPs, 
we will measure such a large invisible decay branching ratio 
of the SM-like Higgs for $m_{\eta} \sim m_H/2$ 
in multi-Higgs doublet models.

As compared to the case of the minimal Higgs portal DM model, 
in the two Higgs doublet portal DM model it is still allowed to have 
a larger value of the invisible decay branching ratio such as $0.8$ (for Set A)  
or even larger.
We emphasize that this conclusion for Set A is almost independent of the type 
of Yukawa interaction, although we have analyzed the invisible decay branching ratio 
assuming the Type-X Yukawa interaction\footnote{See Footnote 3.}.   
Therefore, we conclude that
precise determination of the invisible decay branching ratio
at the LHC or future collider experiments would give
useful information not only for the nature of dark matter
but also for the structure of the Higgs sector
even without detecting any extra scalar boson directly.   \\

\noindent
{\it Acknowledgments}

The work of SK was supported, in part, by Grant-in-Aid for scientific
research (C), Japan Society for the Promotion of Science (JSPS), No.~19540277.

\end{document}